\documentclass[10pt,journal,compsoc]{IEEEtran}

\ifCLASSOPTIONcompsoc
  \usepackage[nocompress]{cite}
\else
  \usepackage{cite}
\fi
 
\ifCLASSINFOpdf
\else
\fi

\usepackage[utf8]{inputenc}
\usepackage{microtype} 
\usepackage{array}
\usepackage{float}
\usepackage{graphicx}
\usepackage[export]{adjustbox}
\usepackage{frame}
\usepackage{tikz}
\usetikzlibrary{calc,shadings,patterns}
\usepackage{hyperref}
\usepackage{hhline}
\usepackage{rotating}
\usepackage{lmodern}
\usepackage{multirow}
\usepackage{multicol}
\usepackage{longtable}
\usepackage{enumitem}
\usepackage{color}
\usepackage[caption=false]{subfig}
\usepackage{booktabs}
\usepackage{pifont}
\usepackage{colortbl}
\usepackage{scalefnt}
\usepackage{pifont}
\usepackage{tabularx}
\usepackage{siunitx} 
\usepackage{crop}
\usepackage[section]{placeins}
\usepackage{comment}
\usepackage{algorithmic}
\usepackage{url}

{}
{}
{}

\begin{document}
\title{A Systematic Identification of Formal and Semi-formal Languages and Techniques for Software-intensive Systems-of-Systems Requirements Modeling}

\author{Cristiane Aparecida~Lana,~\IEEEmembership{}
        Milena~Guessi,~\IEEEmembership{}
        Pablo Oliveira~Antonino,~\IEEEmembership{}
        Dieter~Rombach,~\IEEEmembership{}
        and~Elisa Yumi~Nakagawa~\IEEEmembership{}
\IEEEcompsocitemizethanks{\IEEEcompsocthanksitem C. A. Lana is with the Institute of Mathematics and Computer Sciences (ICMC), University of S\~ao Paulo (USP), S\~ao Carlos 13566-590, Brazil, with the Fraunhofer Institute for Experimental Software Engineering (Fraunhofer IESE), 67663 Kaiserslautern, Germany, and also with Technische Universität Kaiserslautern (TUK), 67663 Kaiserslautern, Germany (cristiane.lana@usp.br; lana@rhrk.uni-kl.de)

M. Guessi is with the Institute of Mathematics and Computer Sciences (ICMC), University of S\~ao Paulo (USP), S\~ao Carlos 13566-590, Brazil (milena@icmc.usp.br)

P. O. Antonino is with the Fraunhofer Institute for Experimental Software Engineering (Fraunhofer IESE), 67663 Kaiserslautern, Germany (pablo.antonino@iese.fraunhofer.de)

Dr. D. Rombach is with the Fraunhofer Institute for Experimental Software Engineering (FraunhoferIESE), 67663 Kaiserslautern, Germany and also with Technische Universität Kaiserslautern (TUK), 67663 Kaiserslautern, Germany (dieter.rombach@iese.fraunhofer.de)

E. Y. Nakagawa is with the Institute of Mathematics and Computer Sciences (ICMC), University of S\~ao Paulo (USP), S\~ao Carlos 13566-590, Brazil (elisa@icmc.usp.br)\protect}
\thanks{This version was submitted to IEEE System Journal on 24 February 2018}}


%
%

\markboth{IEEE Systems Journal, XXXXX Issue XXXX}
{Lana \MakeLowercase{\textit{et al.}}: A Systematic Identification of Formal and Semi-Formal Languages and Techniques for Software-Intensive Systems-of-Systems Requirements Modeling}
%




\IEEEtitleabstractindextext{%
\begin{abstract}
Software-intensive Systems-of-Systems (SoS) refer to an arrangement of managerially and operationally independent systems (i.e., constituent systems), which work collaboratively towards the achievement of global missions. Because some SoS are developed for critical domains, such as healthcare and transportation, there is an increasing need to attain higher quality levels, which often justifies additional costs that can be incurred by adopting formal and semi-formal approaches (i.e., languages and techniques) for modeling requirements. Various approaches have been employed, but a detailed landscape is still missing, and it is not well known whether they are appropriate for addressing the inherent characteristics of SoS. The main contribution of this article is to present this landscape by reporting on the state of the art in SoS requirements modeling.
This landscape was built by means of a systematic mapping and shows formal and semi-formal approaches grouped from model-based to property-oriented ones. Most of them have been tested in safety-critical domains, where formal approaches such as finite state machines are aimed at critical system parts, while semi-formal approaches (e.g., UML and i*) address non-critical parts. Although formal and semi-formal modeling is an essential activity, the quality of SoS requirements does not rely solely on which formalism is used, but also on the availability of supporting tools/mechanisms that enable, for instance, requirements verification along the SoS lifecycle.
\end{abstract}

}

\maketitle

\IEEEdisplaynontitleabstractindextext

%
\IEEEpeerreviewmaketitle

\ifCLASSOPTIONcompsoc
\IEEEraisesectionheading{\section{Introduction}\label{sec:introduction}}
\else
\section{Introduction}
\label{sec:introduction}
\fi

\IEEEPARstart{D}{ue} to the accelerated development of industry and society, several independent software-intensive systems are now able to exchange information and interoperate with each other, resulting in more complex systems, which are called Systems-of-Systems (SoS\footnote{For the sake of simplicity, the acronym SoS is interchangeably used to express singular and plural.}) \cite{Maier:1998, Dahmann:2008, Maier:2014TRM, WhitePaper:2014, Wang:2017QRM}. 
These constituent systems are often developed by different companies and rely on different platforms and technologies. 
The combined work of the constituents allows the SoS to perform complex functions that could not be delivered otherwise \cite{ Maier:1998, Han:2013}.
In this sense, two of the distinguishing characteristics presented by SoS are their evolutionary development and their emergent behavior \cite{Maier:1998, Ncube:2013, WhitePaper:2014, Nielsen:2015}.
\textit{Evolutionary development} refers to the capacity of an SoS to evolve in response to changes in its environment, its constituent systems, or its missions, functions, and purposes.
For instance, an SoS must absorb the constituents' changes, which may affect its mission to resume its proper functioning.
\textit{Emergent behavior} refers to new functions that cannot be realized by any constituent system separately. In fact, these behaviors (or functionalities) can only be realized by the interactions among constituents over time \cite{Keating:2015}.
Emergent behaviors may originate in constituents and trigger new behaviors at the SoS level and vice-versa. 
Examples of SoS can be found in diverse application domains, such as military \cite{BaldwinDod:2008}, aerospace \cite{DeLaurentis;:2011}, transportation \cite{Han:2016SRS}, and health care \cite{Watt:2016}, and are becoming a trend for future systems.

In parallel, Requirements Engineering (RE) is recognized as a fundamental activity of successful software development processes and comprises modeling\footnote{The terms modeling and specification are used interchangeably in this article.}, verification, and validation (V\&V)\footnote{We consider V\&V of requirements as defined in ISO/IEC/IEEE 24765:2017 \cite{ISO24765:2017SSE}.
\textbf{Validation}: confirmation, through the provision of objective evidence, that the requirements for a specific intended use or application have been fulfilled. \textbf{Verification}: the process of ensuring that the software requirements specification complies with the system requirements, conforms to document standards of the requirements phase, and is an adequate basis for the architectural (preliminary) design phase} of requirements. 
For researchers and industry professionals, software projects are extremely vulnerable when RE is conducted poorly \cite{Alshazly:2014}.
In general, 70\% of software projects fail due to low-quality requirements, whose rework cost exceed \$45 billion annually \cite{Randell:2014}.
The cost for correcting errors\footnote{We consider an error as a human action that produces an inconsistent requirements document.} originating in requirements and persisted throughout different development phases can scale up to 100 times of their original costs \cite{Boehm:1981, Maalem:2016}.
To address this risk, RE adopts informal, semi-formal, and formal techniques, methods, and languages (notations) for requirements modeling and V\&V  \cite{Chapurlat:2006}. 
Informal notations are usually more expressive and flexible but they also rely on human expertise \cite{Wang:2016}. 
On the other hand, semi-formal languages\footnote{In the context of this article, UML is a language and each of its diagrams is a technique. Hence, a technique makes it possible to develop a specific structural/behavioral model of a system, while a language expresses systems in a structure that is defined by a consistent set of rules.},
e.g., i*, Unified Modeling Language (UML), and Systems Modeling Language\footnote{SysML is a general-purpose graphical modeling language for representing systems that may include combinations of hardware, software, data, people, facilities, and natural objects \cite{FriedenthalSysML:2008, SysML:2016}} (SysML), provide a defined syntax but lack complete semantics to support communication among the stakeholders \cite{Ahmed:2007}.
Alternatively, formal languages, such as the Vienna Development Method (VDM) and Larch, provide both a well-defined syntax and semantics, being  a set of finite strings of symbols from a finite alphabet \cite{Rozenberg:2004HFL}, but they also require considerably more training than semi-formal languages \cite{Guttag:1993LLT, Wang:2016}. The main advantage of having defined semantics is that it enables the development of automated tools that can efficiently find problems in requirements \cite{Wang:2016}. 
In particular, several formal languages have been tailored to express requirements, such as the Requirements Modeling Language (RML) and the Knowledge Acquisition in Automated Specification (KAOS), which make it possible to precisely state the objectives of software systems.
In this sense, formal languages can support the systematic analysis of formalized statements and their associated impact \cite{Nielsen:2015}, which can be used to reveal missing requirements and inconsistencies, predict behaviors, check for the accuracy of requirements, and also promote the stakeholders' understanding by means of clear semantics \cite{Jarke:2011}.

To deal with requirements related to SoS and constituent systems, RE needs to change its focus to the right composition of constituents within the SoS that yields desirable emergent behaviors at runtime \cite{Ncube:2013}. 
Due to the dynamic nature of SoS, RE is a permanent activity that must be frequently revisited during the SoS life cycle~\cite{Boehm:1984, Bilal:2016}. As a result, SoS have a large and complex set of requirements that have several interdependencies \cite{Easterbrook:1998, Wang:2016}. 
For this reason, informal notations are insufficient for modeling SoS requirements \cite{Wang:2016}. Even though formal techniques can be used for modeling requirements related to SoS autonomy, evolution, and emergent behaviors, these techniques usually rely on approaches that lack support for SoS openness as well as unpredictability \cite{Nielsen:2015}. 
In this scenario, the combination of semi-formal and formal techniques can be sought to balance the limitations of each approach, e.g., using formal techniques to model critical parts of the system and semi-formal ones to model non-critical parts \cite{Ponsard:2005EVV, Zhang:2014MLS}. 
However, a complete landscape about notations and techniques that can be used to model SoS requirements is still missing.

Aiming to address this issue, we present in this article a comprehensive literature review on modeling SoS requirements. 
In particular, we identify which formal and semi-formal languages and techniques have been used for modeling SoS requirements, and how they have been used.
We contextualize the current practice for modeling SoS requirement in regards to specification style and paradigm, and detail known advantages and limitations for many of them, such as UML, SysML, Prototype Verification System (PVS), and Finite State Machine (FSM).
We observe that in spite of several SoS being developed for critical domains, only a few languages and techniques that are used for modeling SoS requirements (e.g., \cite{Atkinson:1991PPS, Ponsard:2005EVV, Yuan:2011MVS, Wang:2017QRM}) are currently supported by automated tools, probably impacting the  accuracy, correctness, and consistency of SoS requirements.

The remainder of this article is organized as follows.
Section~\ref{background} introduces the main concepts related to SoS and further explains the challenges for modeling SoS requirements.
Section~\ref{Relatwoks} provides an overview of related work.
Section~\ref{SM} presents the research protocol for the identification of studies related to the modeling of SoS requirements. Section~\ref{reporting} presents the main results of our review, and Section~\ref{conclusion} concludes this work.

\section{An Overview of Systems-of-Systems and Challenges for Modeling Systems-of-Systems Requirements} \label{background}

SoS are formed of heterogeneous and independent constituent systems that work together to perform a mission \cite{Maier:1998, DeLaurentis:2005, Lane:2013, Maier:2014TRM}.
SoS present a set of inherent characteristics that were originally identified by Maier \cite{Maier:1998} and that have been expanded and rewritten since then to fit specific application domains  \cite{Firesmith:2010, Nielsen:2015}. 
In addition to the two characteristics discussed in Section 1 (i.e., evolutionary development and emergent behavior), there are three others that refer to the properties of constituent systems, namely: (i) \textit{operational independence}, meaning they have their own functionality even when not cooperating with other constituents; 
(ii) \textit{managerial independence}, meaning they are independently managed by their owners; and (iii) \textit{geographical distribution}, meaning they interact among themselves exclusively in terms of information exchange, being distributed over different locations.
Interactions among constituents can lead to the emergence of new SoS behaviors, functionalities, or missions, which might be triggered or influenced by a stimulus from the system's environment \cite{Maier:2014TRM}.
These emergent behaviors enable SoS to provide functionalities that were not originally designed or that cannot be predicted at design time from knowing only what their constituents are \cite{Firesmith:2010, Nielsen:2015}. 

Constituent systems sometimes have unsynchronized lifecycles.
This directly impacts the evolution of an SoS through changes in its fundamental structure, including its constituents and the relationships among them \cite{Han:2013, AxelssonIncose:2015}.
In such cases, the SoS architecture must be open to such changes and evolve accordingly over time in order to embrace new requirements and situations. 

SoS must not restrict the autonomy of constituents, since these constituents can modify their behavior to gain benefits. Moreover, it is necessary to fulfill new requirements and missions of the SoS and also consider the trade-off between constituent missions and SoS missions \cite{AxelssonIncose:2015}.
In this sense, we still need to better understand the principles that control the constituents' behavior in order to align them with the SoS mission \cite{Han:2013}.
Appropriate mechanisms must be identified and put in place to support the fulfillment of SoS missions, including both regulating mechanisms, which minimize inappropriate behavior, and awarding mechanisms, which encourage desirable conduct \cite{Han:2013, Ncube:2013, AxelssonIncose:2015}.

The literature often classifies SoS within four different categories that reflect different levels of managerial control and central authority exercised by the SoS over their constituents as well as different levels of collaboration among such constituents \cite{Maier:1998, Dahmann:2008, BaldwinDod:2008, Madni:2014}.
A Virtual SoS has neither central authority to manage its constituents' activity nor a clear purpose; a Collaborative SoS has its constituents working together more or less voluntarily to fulfill agreed  central purposes; an Acknowledged SoS has its constituents maintaining their independent ownership, objectives, funding, and development approaches, but has no complete authority over its constituents; and a Directed SoS has central management and an engineering team that builds the SoS aiming to fulfill specific purposes whilst having complete authority over the evolution of its constituents. Unlike the latter two types, the first two have no SoS engineering team guiding or managing activities related to the whole SoS.

Because of their inherent characteristics, especially emergent behavior and evolutionary development, the process for requirements modeling for SoS is extremely challenging and should adequately address real-world SoS problems \cite{Walker:2014}. Particular attention should be given to understanding stakeholders' demands, interoperability among constituent systems, architecture, and dynamic evolution, as well as to comprehending how emergent behaviors impact requirements stability \cite{Ncube:2011, Keating:2015}.
Hence, practitioners are sometimes limited in applying traditional approaches (techniques, methods, and tools) to identify the requirements of complex problems in the SoS context.
Approaches should be able to handle several problems associated with the requirements of the whole SoS and, at the same time, should be able to deal with the requirements of the various constituent systems evolvinge independent from each other \cite{Ncube:2011}.
In this sense, the use of new concepts and approaches (such as those based on agents, goals, objectives, and actions) could be explored to make SoS requirements modeling more flexible \cite{Lewis:2009}. 
%


\section{Related Work} \label{Relatwoks}

To the best of our knowledge, there exists no survey, systematic literature review (SLR), or systematic mapping (SM) on formal and semi-formal languages and techniques for modeling SoS
requirements.
Nonetheless, this section discusses two SLRs, one concerning modeling requirements languages \cite{Sepulveda:2016RML}
and the other one concerning specification and methods/techniques for the generation of textual requirements specifications \cite{Nicolas:2009GRS}.
Afterwards, we will also discuss two works on the comparison of specific formal languages for the description of requirements \cite{ Sharma:2013CFS}.

Selp\'uvida et al. \cite{Sepulveda:2016RML} analyzed 54 studies to evaluate requirements modeling languages in terms of their maturity, level of expressiveness, origin (i.e., developed in industry, academia, or both), context of use (i.e., industry or academia), and validation. Their work was developed in the context of software product lines. As their main result, the authors point out the lack of a conceptual foundation and tool support for using these languages.
N\'icolas and Toval \cite{Nicolas:2009GRS} evaluated 30 studies, also in the context of software product lines, and report several methods and techniques that can support the generation of textual requirements specifications from software models (i.e., graphical models). The authors mention tool support as a concern and noticed several difficulties in finding the business requirements and the rationale underlying graphical models.
Their work does not classify such methods and techniques with regards to formal or semi-formal issues.

Dutertre and Stavridou \cite{Dutertre:1998ASR} present a terminology and a comparative analysis between two formal specification languages, namely Requirements State Machine Language (RSML) and Software Cost Reduction (SCR). 
Their work compares these languages in the avionics domain with the aim of improvinge requirements traceability. Thus the focus is not on narrow language characteristics, but rather on performance.
Although the authors present a detailed analysis together with information about traceability performance, the work is restricted to these two languages and one application domain, while our SM is more comprehensive and also investigates the combination of languages for requirements modeling, in addition to modeling techniques.
Similarly, Sharma and Sing \cite{Sharma:2013CFS} compare the syntax of five formal languages, namely Z, Object Constraint Language (OCL), VDM, Specification and Description Language (SDL), and Larch.
The goal of this work at comparison using parameter specification types and mathematical types used to support the choice of an appropriate language for a particular problem.

Finally, we found that all related work focuses on more specific issues, while our work is a wider investigation, analyzing semi-formal and formal ways to model SoS requirements.

\section{Planning and Conduction of the Systematic Mapping} \label{SM}

Our SM was conducted from April 2017 to August 2017 by software engineering researchers and requirements engineering experts.
To conduct this mapping, we followed the systematic process proposed by Petersen et al. \cite{Petersen:2015GCS} and Kitchenham and Charters \cite{Kitchenham:2007}.
In short, this process is composed of three main phases (planning\footnote{The complete research protocol can be found at \url{https://goo.gl/yntV3m}.}, conduction, and reporting), which will be explained in more detail in the following sections.

\subsection{Planning} \label{planning}

Aiming to find all relevant primary studies for our research,if possible, we established the following research question (RQ):\\


\textit{RQ: Which formal and semi-formal languages and techniques have been used for modeling SoS requirements?} \\

The search strategy for answering this RQ is based on two main keywords, namely ``requirement modeling'' and ``formal\footnote{The keyword ``semi-formal'' was not included in the search string, because our string already includes the term ``formal'', which allows finding studies related to semi-formal languages/techniques. This was confirmed in our pilot study.}''.
To avoid missing out on any relevant study, we also considered the keywords ``requirement verification'' and ``requirement validation'' for our search string. Additional terms were also identified together with the opinion of experts in RE and SoS. After a number of refinement iterations and string calibration, the final search string used in this SM was:\\

\small{\textsf{(``requirement modeling''  OR  ``model of requirement''  OR  ``requirement model''  OR  ``modeling requirement''  OR  ``modeling of requirement''  OR  ``requirement representation''  OR  ``requirement analysis''  OR  ``analysis of requirement''  OR  ``requirement analyzing''  OR  ``requirement design''  OR  ``requirement verification''  OR  ``verification of requirement''  OR  ``evolution of requirement''  OR  ``requirement evolution''  OR  ``requirement validation''  OR  ``validating requirement''  OR  ``validation of requirement''  OR  ``requirement specification''  OR  ``quality requirement''  OR  ``non functional requirement''  OR  ``non-functional requirement''  OR  ``nonfunctional requirement''  OR  ``non functional property''  OR  ``non-functional property''  OR  ``nonfunctional property''  OR  ``non functional characteristic''  OR  ``non-functional characteristic''  OR  ``nonfunctional characteristic''  OR  ``quality attribute''  OR  ``quality characteristic''  OR  ``quality factor''  OR  ``quality criterion'' )  AND  ( formal ))}}. \\

\normalsize It is worth highlighting that the term ``system-of-systems'' (and its synonyms such as SoS) was not included in our string because this term is not necessarily used by all studies that address SoS. Besides that, other related terms, such as ``cyber-physical system'', ``ecosystem'', and ``distributed system'', could be used to refer to an SoS. Therefore, neither ``system-of-systems'' nor related terms were included in the string. Hence, the selection of studies that address SoS was conducted by reading each study in its entirety. 

To select publication databases for our mapping, we followed specific criteria \cite{Dieste:2009, Kitchenham:2007}. 
The ACM Digital Library, ISI Web of Science, IEEE Xplore, Science Direct, Scopus, and SpringLink databases are the most relevant ones for computer science, and are widely used in software engineering \cite{Pertensen:2015, Kitchenham:2010, munir:2014}.
Our SM was also complemented by manual selection of studies from publication venues that are not indexed by any of these databases.


\subsubsection{Inclusion and Exclusion Criteria}

We defined specific selection criteria for evaluating primary studies recovered from publication databases. These criteria were applied in the first and second round of selection to identify relevant studies.
To include a study in our mapping, we had only one criterion: studies that address formal and/or semi-formal modeling of SoS requirements and similar systems (that have the characteristics of SoS).
On the other hand, we had several criteria for excluding a study from our SM. A study was excluded if it is related to the modeling
of monolithic systems or if it is an editorial, keynote, opinion, tutorial, panel, extended abstract\footnote{For this work, we considered as extended abstracts all studies with up to three pages.}, or gray literature (e.g., technical report). 
We also excluded studies where a newer or more complete version exists. For example, study \cite{Krishna:2009CPC} is more complete than study \cite{Krishna:2004ACS}, hence only the former is included in our mapping. Besides, studies in languages other than English and those whose full text is not available were also excluded.


\subsubsection{Data Extraction and Synthesis Method}

The data extracted from each primary study was managed with the support of the Parsif.al\footnote{\url{https://parsif.al/}} and Tableau Public\footnote{\url{https://public.tableau.com/s/}} tools, and MS Excel.
Data extraction was accomplished by the first author of this article and discussed and reviewed by the other authors. Concordance meetings were also conducted when necessary to discuss data and their relationships. To summarize and present these data, we used qualitative analysis methods.


\subsection{Conduction} \label{conduction}

The primary studies were selected following the protocol briefly described in Section \ref{planning}.
4,751 studies were obtained: 3,993 unique studies were recovered from six publication databases and 758 from manual selection in publication venues. 
After removing duplicate studies (i.e., 839 of them), 3,912 studies remained for selection.
Initially, the title, abstract, and, when necessary, the introduction section of each study were read and the selection criteria were applied.
In this way, a total of 96 studies were selected.
The full text of each study was then read and the selection criteria were applied again.
As a result, a set of 25 primary studies were selected for data extraction.
Besides, one study \cite{Zhang:2012AOF} was inserted following the suggestion of an expert, bringing the total to 26 studies. Table \ref{tab:includedstudies} presents these studies, their ID, author name (s), publication title that are external link where their study was published, and publication year.

\begin{table*}[htb]
\centering
\caption{Primary studies included}
\vspace{0.1cm}
\label{tab:includedstudies}
\begin{tabular}{|p{0.4cm}|p{4.8cm}|p{10.4cm}|p{0.6cm}|}
\hline
\textbf{ID} & \textbf{Author}    &   \textbf{Title}                                                    & \textbf{Year} \\
\hline \hline
S1          & Atkinson, W. and Cunningham, J. &     \href{http://ieeexplore.ieee.org/document/73716/}{Proving properties of a safety-critical systems}                      & 1991          \\ \hline
S2          & Leveson, N.G. and Reese, J.D.      &  \href{http://ieeexplore.ieee.org/document/317428}{Requirements specification for process-control systems}            & 1994        \\ \hline
S3          & Heimdahl, M.P.E. and Leveson, N.G. &  \href{http://ieeexplore.ieee.org/document/508311/}{Completeness and consistency in hierarchical state-based requirements}              & 1996           \\ \hline
S4          & Crow, J. and Di Vito, B.           &  \href{https://dl.acm.org/citation.cfm?id=287023}{Formalizing Space Shuttle software requirements: four case studies}                & 1998       \\ \hline
S5          & Jong, E. et al.                         & \href{http://ieeexplore.ieee.org/document/839888/}{Refinement in requirements specification and analysis: a case study}               & 2000        \\ \hline
S6         & S\'anchez-Alonso, M. and Murillo, J. M. & \href{http://wer.inf.puc-rio.br/WERpapers/artigos/artigos\_WER02/alonson.pdf}{Specifying cooperation environment requirements using formal and graphical techniques}     & 2002  \\ \hline
S7         & Ponsard, C. et al.           &      \href{http://www.sciencedirect.com/science/article/pii/S1571066105050334}{Early verification and validation of mission critical systems}                         & 2005           \\ \hline
S8         & Linhares, M. V. et al.   &    \href{http://ieeexplore.ieee.org/document/4416788/}{Introducing the modeling and verification process in SysML}                                & 2007              \\ \hline
S9        & Ghazel, M. and El Koursi, E.M.  & \href{http://ieeexplore.ieee.org/document/4304240/}{Automatic level crossings: from informal functional requirements' specifications to the control model design}    & 2007   \\ \hline
S10         & Jamal, M. and Zafar, N.A.    &  \href{http://ieeexplore.ieee.org/document/4381340/}{Requirements analysis of air traffic control system using formal methods}                  & 2007              \\ \hline
S11         & Goldsby, H. J. et al.    & \href{http://ieeexplore.ieee.org/document/4492385/}{Goal-based modeling of dynamically adaptive system requirements}                              & 2008                \\ \hline
S12         & Krishna, A. et al.       & \href{http://ieeexplore.ieee.org/document/4492385/}{Consistency preserving co-evolution of formal specifications and agent-oriented conceptual models}   & 2009           \\ \hline
S13         & Sun, H. et al.           & \href{http://ieeexplore.ieee.org/document/5635065/}{Automata-based verification of security requirements of composite Web Services}                 & 2010                 \\ \hline
S14         & Tang, W. et al.              & \href{http://ieeexplore.ieee.org/document/5486079/}{Scenario-based modeling and verification for CTCS-3 system requirement specification}         & 2010               \\ \hline
S15         & Tang, W. et al.           &  \href{https://www.witpress.com/Secure/elibrary/papers/CR10/CR10069FU1.pdf}{Scenario-based modeling and verification of system requirement specification for the European Train Control System} & 2010    \\ \hline
S16         & Whittle, J. et al.    &  \href{https://link.springer.com/article/10.1007/s00766-010-0101-0}{RELAX: A language to address uncertainty in self-adaptive systems requirement}                 & 2010       \\ \hline
S17         & Yuan, L.  et al.   & \href{http://ieeexplore.ieee.org/document/5741283/}{Modelling and verification of the system requirement specification of train control system using SDL}    & 2011     \\ \hline
S18         & Cimatti, A. et al.  & \href{https://link.springer.com/article/10.1007/s10270-009-0130-7}{Formalizing requirements with object models and temporal constraints}                           & 2011               \\ \hline
S19         & Zhang, L.  & \href{https://ai2-s2-pdfs.s3.amazonaws.com/f558/42d4097e271e901622e5fed6ed1d6d22bfa0.pdf}{Aspect-oriented formal techniques of cyber physical systems}                                     & 2012                     \\ \hline
S20         & Deb, N. and Chaki, N.  & \href{http://ieeexplore.ieee.org/document/6785520/}{Verification of i* models for existential compliance rules in remote healthcare systems}         & 2014         \\ \hline
S21         & Zhang, L.   & \href{http://ieeexplore.ieee.org/document/6935460/}{Modeling large scale complex cyber physical control systems based on system of systems engineering approach}     & 2014     \\ \hline
S22         & Zou, L. et al.   & \href{https://link.springer.com/chapter/10.1007/978-3-642-54108-7\_14}{Verifying Chinese train control system under a combined scenario by theorem proving}             & 2014                 \\ \hline
S23         & Chen, Z. et al.   & \href{http://www.sciencedirect.com/science/article/pii/S2351978915002358}{Exploring a timed-automata fuzzy cognitive maps based approach for modeling systems of systems}  & 2015                 \\ \hline
S24         & Piccolo, A. et al.   &  \href{http://ieeexplore.ieee.org/document/7101503/}{Use of formal languages to represent the ERTMS/ETCS system requirements specifications}           & 2015           \\ \hline
S25         & Han, L. et al.   & \href{http://ieeexplore.ieee.org/document/7552030/}{Safety requirements specification and verification for railway interlocking systems}              & 2016                \\ \hline
S26         & Wang, Q.-L. et al. & \href{https://link.springer.com/article/10.1631/FITEE.1500309}{A quality requirements model and verification approach for system of systems based on description logic}  & 2017          \\ 
\hline
\end{tabular}
\end{table*}

\subsection{Threats to Validity}\label{sec:threatsToValidity}

To increase the trustworthiness of our SM and minimize biases that could be introduced by the authors \cite{Wohlin:2012}, we identified potential threats to validity and discuss below the actions we put in place to mitigate them.

\begin{itemize}[leftmargin=0.4cm]

    \item \textit{Missing important primary studies:}
    As already considered in Kitchenham and Charters' guidelines \cite{Kitchenham:2007}, it is not possible to guarantee the identification of all relevant primary studies that exist in the literature for a given research topic. 
    To mitigate this threat, we carefully established and validated the research protocol together with experts. We also conducted a pilot study and defined the selection criteria in order to minimize the risk of excluding relevant studies.
    Besides that, we adopted a specific group of databases considered as the most relevant ones  for computer science according to \cite{Pertensen:2015, Kitchenham:2010, munir:2014}.
    However, primary studies indexed by other databases 
    could be missing. 
    To mitigate this threat, we also performed a manual search of publication venues and studies recommended by experts consulted for this mapping.
   
    \item \textit{Selection of primary studies:}
    Studies were selected based on how the authors referred to their studies, i.e., modeling, specification, validation, and verification. However, there is a narrow line between these terms/concepts when considering RE. Indeed, in practice, these terms have been used interchangeably and even the context of their application does not make the meaning clear.
    So, whenever there was doubt about the inclusion of any primary study, it was discussed with experts during the concordance meetings. 
   
    \item \textit{Number of reviewers and reliability:}
    Our SM was conducted by three researchers.
    This number of reviewers may imply some risk of bias.
    To ensure the reliability of our SM and to obtain an unbiased selection process, our planning phase, including the research question and the selection criteria (i.e., inclusion criterion and exclusion criteria), was carefully set up in advance.
    The research question and the selection criteria are detailed enough to make it possible to reproduce the steps to obtain the 26 primary studies selected. 
    In addition, SoS and RE experts supported the entire selection process through concordance meetings. 
     
     \item \textit{Non-available studies and data extraction:}
    88 primary studies obtained in the databases (i.e., 2.2\% or 88/3,993) were not available.
    Despite efforts to contact the authors by email or via social network such as ResearchGate\footnote{\url{https://www.researchgate.net/}}, we did not gain access to them.
    Moreover, some information described in the included studies was not clear and had to be interpreted.
    To ensure the validity of our SM, discussions with experts were carried out whenever there were doubts.

\end{itemize}

\section{Results} \label{reporting}

The following sections report the findings of our SM. First, Section \ref{general_results} presents an overview of primary studies that were used for answering our RQ in Section \ref{answers_rq1}. Our main findings are then discussed in Section \ref{discuss}.

\subsection{General Results}\label{general_results}

\begin{table*}[htb]
    \caption{Characterization of the included primary studies}
    \label{tab:ComparativeTable}
    \centering
    \includegraphics[width=.83\linewidth]{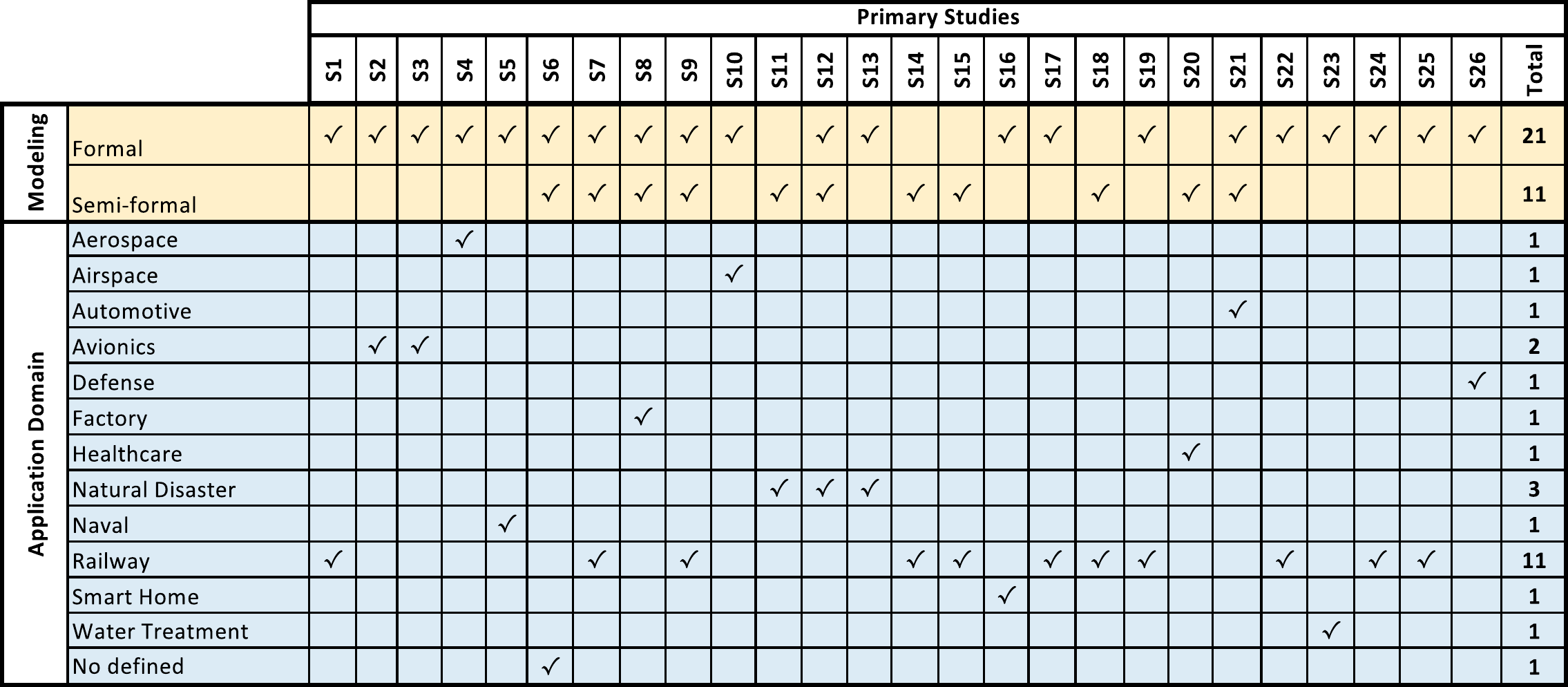}
\end{table*}

Table \ref{tab:ComparativeTable} shows an overview of the 26 primary studies included in our SM. Taking into the account challenges for modeling SoS requirements (cf. Section \ref{background}), we can already observe several initiatives that use formal and/or semi-formal languages for addressing these issues. In fact, 80.8\% of the included studies model SoS requirements by means of formal languages or techniques.
This choice can be justified by the critical application domains in which these studies were performed and also by regulations such as DO278 \cite{EUROCAE:2002GCN} and ISO/DIS 26262 \cite{ISO26262:2009ROAD}, which require specific procedures for the certification of these systems.

Our SM revealed the use of semi-formal languages and techniques in eleven studies, of which six apply a mix of formal and semi-formal languages and techniques (i.e., S6, S7, S8, S9, S12, and S21). Formal languages and techniques are specifically used in these studies for modeling critical parts of the system (e.g., brakes or rudder control systems of an airplane) whilst non-critical parts have been modeled using semi-formal languages \cite{EUROCAE:2002GCN, NasaCofer:2014DO333}.
The remaining five studies (i.e., 5 of 11) only used semi-formal languages (i.e., S11 and S20) or semi-formal techniques (i.e., S14, S15, and S18) for modeling SoS requirements.
For instance, S14 \cite{Tang2010CTCS3} and S15 \cite{Tang:2010EuroTC} apply UML for modeling scenario-based requirements of a train control system. In particular, the authors use Sequence Diagrams (UML-SD) for the representation of interactions and behaviors of this system and its components  exhibited in operations.

Overall, the 26 studies included in this SM cover 12 different critical application domains, as depicted in Table \ref{tab:ComparativeTable}. The most thoroughly explored domain is railway systems, which is addressed by 42.31\% (11 studies) of the included studies, followed by natural disaster (3 studies) and avionics (2 studies).
Only one included study, S6, does not focus on any particular application domain. In S6, the authors investigated formal and graphic techniques for complex systems in general, but evaluated their work in the context of the fire control system for a museum. 
All domains identified in this SM are safety-critical, i.e., system failures can harm people and/or the environment, or result in significant financial loss. 
As SoS have become increasingly predominant in safety-critical domains, as observed in this SM, more rigor is certainly needed to properly deal with requirements, which is reflected in the greater interest in formal approaches.


\subsection{RQ: Formal and Semi-formal Languages and Techniques for SoS Requirements Modeling}\label{answers_rq1}

\begin{figure*}[!ht]
 	\centering
 	\includegraphics[width=\linewidth]{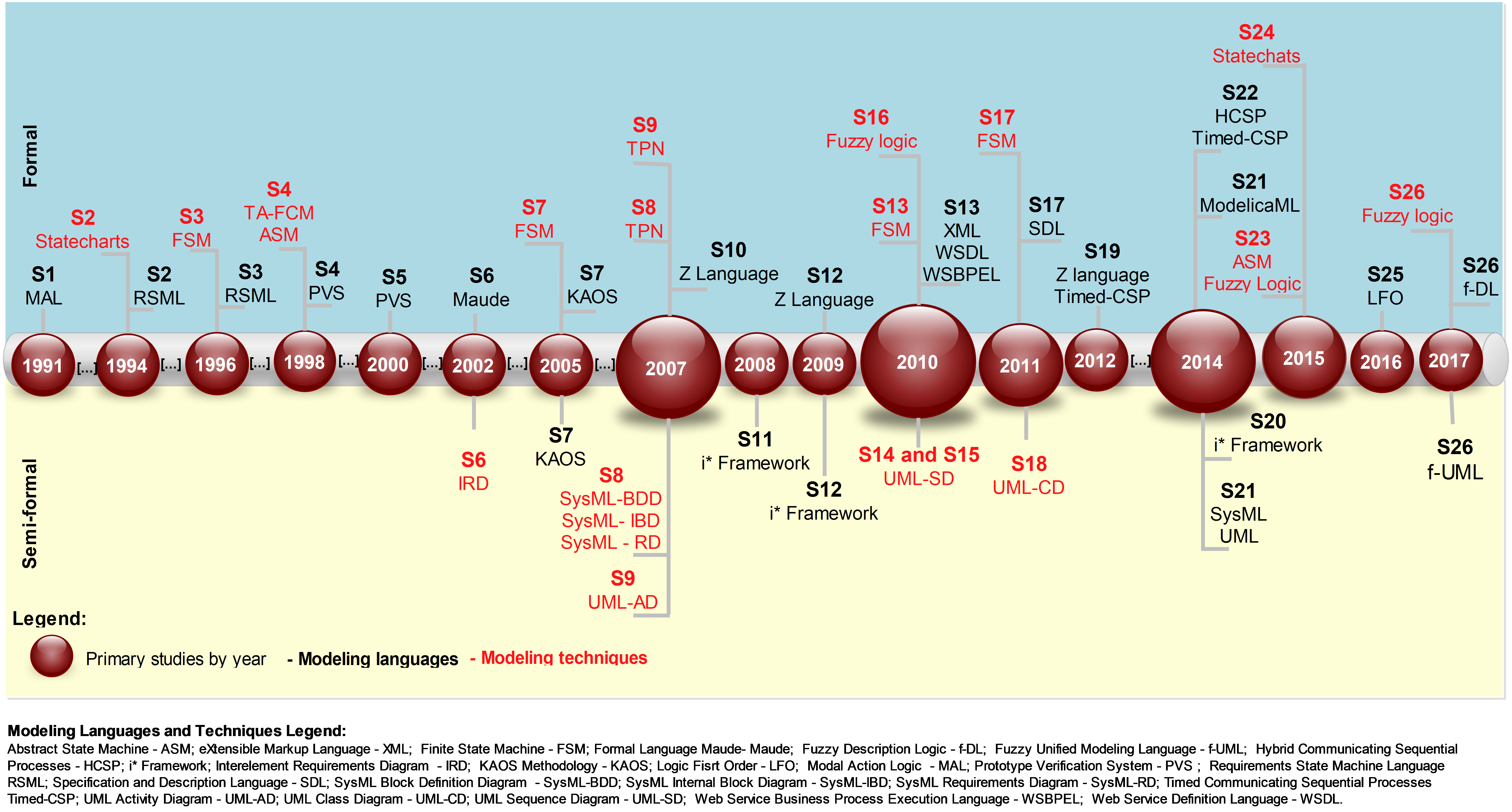}
 	\caption{Overview of the formal and semi-formal techniques and languages per year}
 	\label{fig:OverviewperYear}
 \end{figure*}

Different formal and semi-formal languages and techniques have been adopted to model SoS requirements. Some identified languages are variants of others, e.g., Timed-CSP is a variant of the parallel language Communication Sequential Process (CSP).
In particular, we identified 19 languages in the 26 included studies, of which 15 are formal and five are semi-formal. We also identified 13 techniques, of which six support formal modeling and seven support semi-formal modeling. These languages and techniques are represented in Figure \ref{fig:OverviewperYear}.
In this figure, number of studies included per year is reflected in the size of the circles that represent years. 
For instance, the year 2010 is represented by a larger circle than the year 1991 since it contains four studies, whereas the latter contains only one.
Formal languages and techniques are represented in the top (blue) region of this figure, while semi-formal languages and techniques are shown in the bottom (yellow) region. We also distinguished modeling languages, which are shown in black font, from modeling techniques, which are shown in red font. For instance, the year 2010 contains two studies (i.e., S14 and S15) that use semi-formal languages (i.e., UML-SD), one study (i.e., S13) that uses a formal language, and two studies (i.e., S13 and S16) that use formal techniques. In particular, S13 combines three formal languages, namely eXtensible Markup Language (XML), Web Service Definition Language (WSDL), and Web Service Business Process Execution Language (WS-BPEL), in addition to using the FSM technique, and finally, S16 adopted only Fuzzy Logic. 

\begin{figure}[!ht]
 	\centering
 	\includegraphics[width=\linewidth]{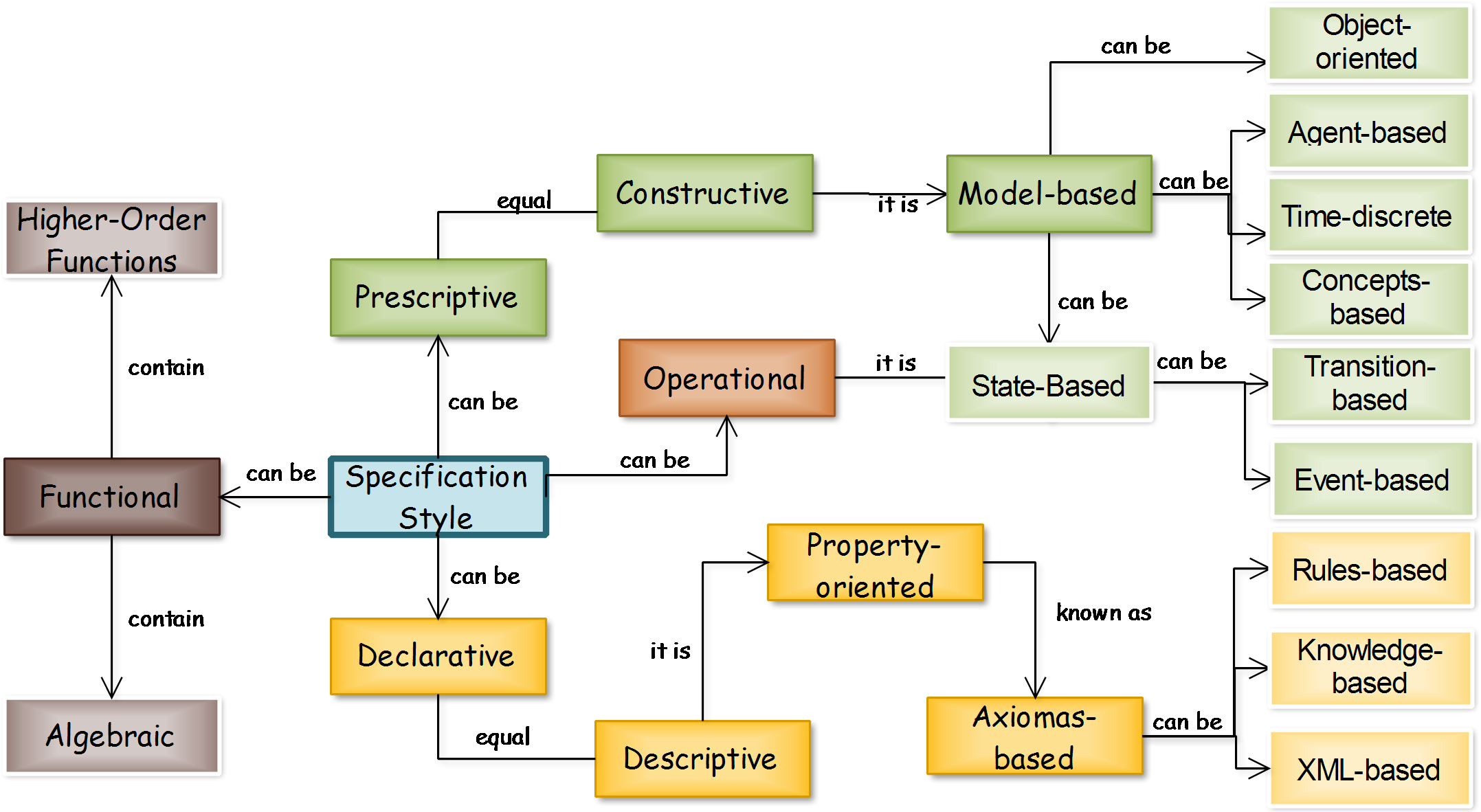}
 	\caption{Conceptual model of specification style terminologies}
 	\label{fig:terminology}
 \end{figure}

Formal and semi-formal languages and techniques can be classified and compared in terms of specification styles, paradigm, and executable syntax. This comparison was done according to the specification styles described in \cite{Misic:1997FSS, Nasa:1997FMS, NasaSrivas:1995FVA, Lamsweerde:2000FSR, Ghezzi:2002}.
The most important characteristic of the specification language is based on its mathematical foundation \cite{Misic:1997FSS}. 
However, different terminologies are used in each study for the same style. For example, the technical report of NASA \cite{Nasa:1997FMS, NasaSrivas:1995FVA} classifies languages and techniques as either model-oriented or property-oriented. The former can be considered as a constructive (or prescriptive) style, whereas the latter can be considered as a declarative (or descriptive) style, which is also referred to as axiom-based or rule-based by other studies \cite{Lamsweerde:2000FSR, Ghezzi:2002}.
To clarify the classification of specification styles used in this SM, we created the conceptual model presented in Figure \ref{fig:terminology}. We refer to this conceptual model in the column ``Specification Style'' in Table \ref{tab:ClassificationStyle}.

\begin{sidewaystable*}
    \caption{Classification of formal and semi-formal languages and techniques}
    \label{tab:ClassificationStyle}
    \centering
    \includegraphics[width=\linewidth]{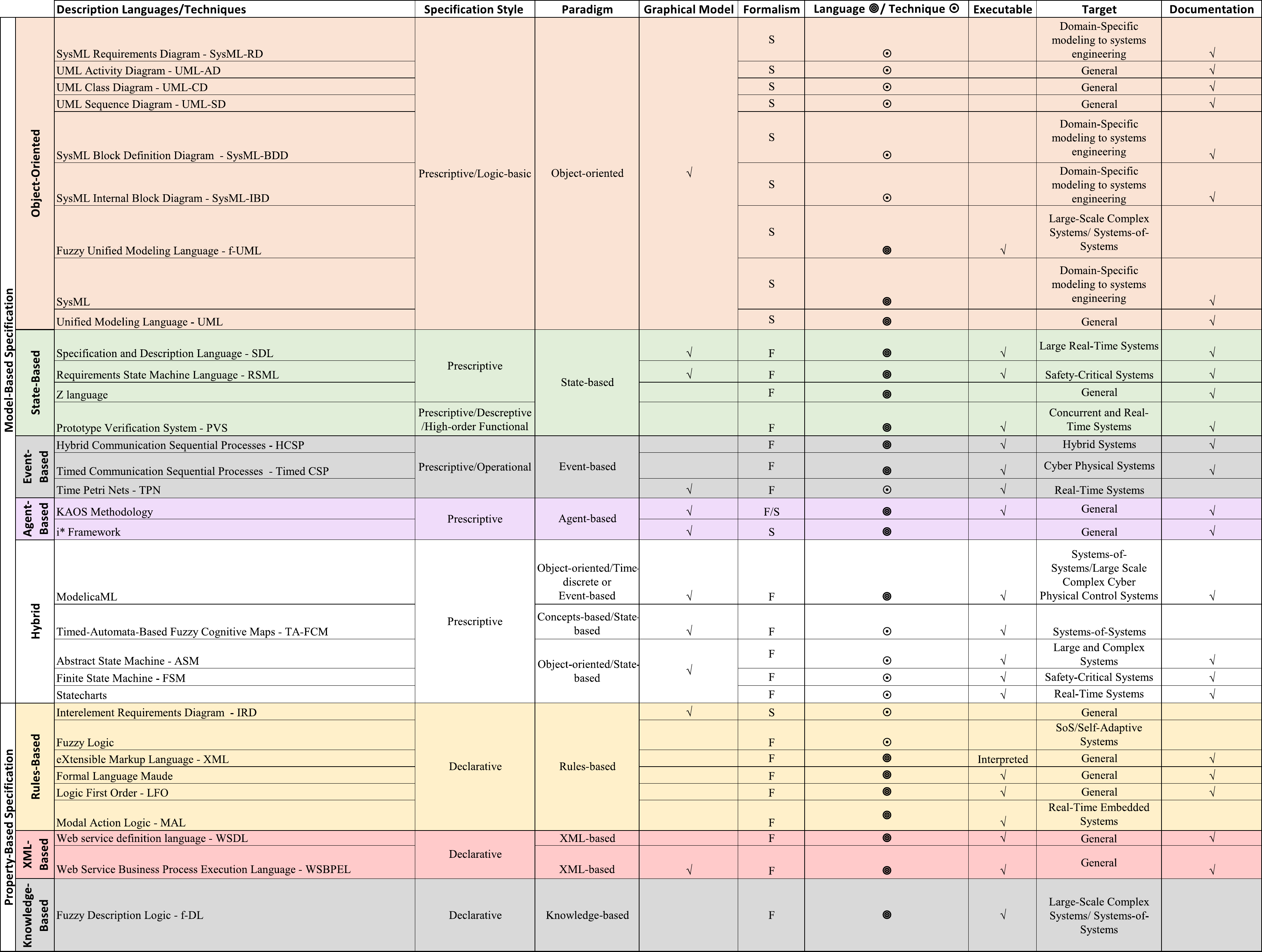}
\end{sidewaystable*}

A model-based specification, also known as constructive and prescriptive style, describes the desired behavior for an intended system usually by means of models \cite{Ghezzi:2002}. While this specification style is often able to create detailed models of the system, an excess of information can also lead to bias in design and implementation \cite{NasaSrivas:1995FVA}. Nonetheless, model-based specifications are often easier for non-technical users to understand than property-oriented specifications. The latter can be used to describe desired properties of a system at a higher abstraction level, which generally leads to a specification with fewer details \cite{Nasa:1997FMS, Ghezzi:2002}. On the other hand, it is easier to introduce inconsistencies in this style, which also requires advanced knowledge to be read and understood by users \cite{Nasa:1997FMS, NasaSrivas:1995FVA, Ghezzi:2002}. 
A detailed discussion on the trade-offs between the two specification styles can be found in \cite{Nasa:1997FMS, NasaSrivas:1995FVA}.

Other characteristics analyzed in Table \ref{tab:ClassificationStyle} include:
\begin{itemize} [leftmargin=0.4cm]
\item \textit{Paradigm}: classifies a language or technique according to modeling features; 
\item \textit{Graphical Model}: indicates if a language or technique supports a graphical notation;
\item \textit{Formalism}: indicates whether a language or technique is formal (F) or semi-formal (S); 
\item \textit{Language/Technique}: indicates whether it is a language or technique;
\item \textit{Executable}: indicates whether is possible to execute the models created using the language/technique for the purpose of requirements V\&V, which could minimize time and costs \cite{Sammi:2010FSL};
\item \textit{Target}: indicates which sorts of systems are described using the language or technique (e.g., large real-time system or systems-of-systems); and
\item \textit{Documentation}: indicates the existence of external information for the language or technique. For example, f-UML is only described in the included studies whereas the Z language is also described by the ISO/IEC13568:2002 standard \cite{ISO13568:2002Zlang}.
\end{itemize}

\subsubsection{Model-based Specification}

Our SM identified five model-based specification paradigms: object-oriented, state-based, event-based, agent-based (also known as goal-based), and hybrid\footnote{We classify as a hybrid paradigm those languages and techniques that can be classified in more than one paradigm.}.
The hybrid paradigm encompasses one language, i.e., ModelicaML, and four techniques, i.e., statecharts, FSM,
abstract state machines, and Timed-Automata-based Fuzzy Cognitive Maps (TA-FCM).
The object-oriented paradigm has a great impact on systems development and, consequently, on requirements modeling. Our SM identified the use of UML and its variants such as SysML. For instance, 
S21 combines UML and SysML to ModelicaML, which is deriveds from UML and SysML, to model the requirements of a cyber physical SoS. 
In particular, S8 combines the Requirements Diagram, the Block Definition Diagram, and the Internal Block Diagram of SysML with Time Petri Nets for modeling and model checking the behavior requirements of a factory plant. 
UML techniques, specifically Activity Diagram, Class Diagram, and Sequence Diagram, are used for representing requirements in S8, S8, S14, S15, and S18. 
Even though our SM identified UML and SysML as recurrent languages for the development of object-oriented systems in industry, these languages and their variants such as executable UML (xUML\footnote{\url{https://xtuml.org/}}) still lack formal execution semantics that can support formal modeling. For this reason, we have also noticed their combination with formal approaches~\cite{Haimes:2012}.


Most included studies combine two or more languages and techniques for modeling SoS requirements. Particularly, S2, S3, S4, and S17 combine formal languages and formal techniques; S7 and S12 combine formal languages and semi-formal techniques; S8 and S14 combine formal and semi-formal techniques; and S19, S21 and S22 combine formal and semi-formal languages.
For instance, S17 employs one formal language, i.e., the Specification and Description Language (SDL), and one formal technique, i.e., Finite State Machines (FSM), in a method for the modeling and verification of the systems requirements of a train control system.
This leads us to infer that SoS requirements modeling is not completed using a single formal language or technique but instead requires a set of languages and techniques to properly address the modeling challenges posed by each domain. 
In this sense, more studies are needed to investigate SoS requirements modeling in other domains, considering the different challenges that might be faced in each of them, and to evaluate which formal and/or semi-formal languages and techniques provide better support for SoS inherent characteristics.


Our SM identified two languages that support agent-based paradigms in industrial environments \cite{Goldsby:2008GMD, Piccolo:2015UFL}, namely the i* framework and the KAOS methodology.
Agents can be defined as active components representing people, devices, legacy software, or software-to-be, which are responsible for fulfilling specific requirements and expectations \cite{RespectKAOS:2007}.
The i* framework is a conceptual modeling language adopted by S11, S12, and S20. 
This language is used for modeling the environment of a system-to-be, supporting critical modeling decisions, such as identification of the main system's goals, representation of multiple stakeholders and their interdependence, and possibilities for exploring the use of these relationships \cite{Krishna:2009CPC, YuStar:2011}.
These characteristics can be interesting for SoS requirements modeling since SoS are formed by interacting independent constituents that individually address a particular set of stakeholders and goals \cite{Bendov:2009,Maier:1998, Han:2013}.
Aside from agents, the KAOS methodology adopted by S7 represents goals that express desired system properties stated by stakeholders that are met by agents \cite{RespectKAOS:2007}. In general, the language supports four techniques: 
(i) goal model: forms a set of interrelated goal diagrams that have been put together to address a particular problem; 
(ii) object model: describes objects (e.g., agents, entities, and relationships); 
(iii) responsibility model: it describes which requirements and expectations an agent is assigned to; and 
(iv) operation model: describes all behaviors that agents need to perform to fulfill their requirements \cite{Ponsard:2005EVV, RespectKAOS:2007}. 
KAOS assists in the establishment of both formal and semi-formal models. Semi-formal models are based on text and include graphical representations, whereas formal models are built on top of semi-formal models, either partially or entirely \cite{RespectKAOS:2007}.

The state-based paradigm can be considered in the behavioral paradigm, since it is used to represent systems behavior by transitioning among different states. In critical systems, models depicting behavior diagrams such as FSM 
and sequence diagrams, have been employed not only in the modeling of requirements for safety-critical systems but also in their verification by means of simulations.
Simulations are traditionally used for evaluating different execution scenarios at design time and, more recently, have also been used for simulating SoS \cite{Honour:2013, Neto:2014}. However, SoS simulation is complicated due to a combination of factors, such as performance issues, conflicting goals, standards, and emergent behaviors (which might be known or unknown at design time) \cite{Honour:2013, Zeigler:2015}. In spite of this difficulty, simulation also brings important benefits such as early identification of errors and problems that can be corrected before the actual realization of the SoS \cite{Neto:2014}.
Languages and techniques supporting the state-based paradigm were used by 52\% of the included studies (i.e., S2, S3, S4, S5, S7, S10, S12, S13, S17, S19, S23, S24, and S26).
An example is S8, which used PVS, a language with tool and theorem proof integrated with decision procedures for different theories including real and integer arithmetics, in conjunction with an abstract state machine and TA-FCM. 
This study investigates the formalization of subsystem modeling of NASA's Space Shuttle using PVS to explore and document the feasibility and utility of formalizing critical Shuttle software requirements representing a spectrum of maturity levels. PVS is still being used at NASA Langley Research Center (LaRC) for modeling requirements of aerospace applications, such as the pilot flying specification \cite{NasaCofer:2014DO333} and the aerospace verification tool \cite{NasaWagner:2017Tool}.

Although we have classified PVS as a model-based style (i.e., state-based paradigm), it can also be considered a property-based style (i.e., axiom-based paradigm), since a set of properties are described to ensure consistency of the specification \cite{NasaSrivas:1995FVA, Nasa:1997FMS}. In addition, it also considers higher-order logic or higher-order function, which is classified as a functional style \cite{Lamsweerde:2000FSR}. The same author classifies Time Petri Net as an operational style, while others classify it as a model-based style \cite{NasaSrivas:1995FVA, Nasa:1997FMS, Ghezzi:2002}.

\subsubsection{Property-based Specification}
Our SM also identified three property-based paradigms: 
(i) the rule-based paradigm, which uses formal and semi-formal languages and techniques; 
(ii) the XML-based paradigm, which uses formal languages; and 
(iii) the knowledge-based paradigm, which uses formal languages.
Even though there are fewer property-based paradigms, they have been used in SoS requirements modeling \cite{Cordes:1988GRS, Nguyen:2014, abdalla:2015, NasaCofer:2014DO333}.
All identified property-based languages, namely XML, WSDL, WS-BPEL, FOL, Maude, Modal Action Logic (MAL), and fuzzy Description Logic (f-DL), and property-based techniques, namely Interelement Requirements Diagram (IRD) and Fuzzy Logic, were sometimes applied separately by each identified study. 
For instance, S26 adopted the Description Logic ontology for the description of the quality requirements, which is a logical reconstruction of a frame-based knowledge representation language.
To specify functional and non-functional SoS requirements that are considered fuzzy and vague requirements at the mission level an extension of f-DL called f-SHIN was applied to describe the necessary quality of the requirements, as it has a strong ability of representation and decidability. 
f-DL is introduced to formalize the UML model and to provide an algorithm for converting a fuzzy UML (f-UML) model into the f-DL ontology to automate verification.

With regard to SoS requirements verification, S25 focus on automation and verification of safety requirements based on pattern-based specification. 
These requirements were specified using First Logical Order (FOL), which belongs to the rule-based paradigms, to verify safety requirements in the railway domain automatically through model checking. Due to the characteristics of SoS, the creation of an environment for verifying requirements is challenging. 
Each constituent has its own verification responsibility; therefore changes in constituents may result in verification events that might affect the entire SoS \cite{Honour:2013}. 
Moreover, the managerial independence of the constituents often does not allow synchronization of multiple life cycles. Hence, verification of SoS requirements and of the SoS itself sometimes accurs without the presence of all the constituents participatinge in an SoS (i.e., without all capabilities) \cite{Lewis:2009, Luna:2013}. Besides that, due to the evolutionary development of SoS, requirements modeling and verification should be performed continually, including evaluation of the system's capability regarding its missions \cite{Dahmann:2010,Honour:2013}.


\subsection{Discussion} \label{discuss}


Our investigations suggest that formal and semi-formal modeling of SoS requirements is an essential activity for developing these systems that requires more rigor in the requirements specification process. Hence, applying more formal approaches in SoS modeling and V\&V can increase the consistency, correctness, and completeness of their requirements.
Similarly, industrial standards have also reinforced this necessity, mainly by means of certification of these systems. Example as, the DO278 \cite{EUROCAE:2002GCN} used in the certification of avionics, IEC/TR 80002 \cite{AAMI:2009MDS} used in medical device development, and EN 50126 \cite{EN50126:1999RAMS},ISO/DIS 26262 \cite{ISO26262:2009ROAD} applied in road vehicles, and EN 50128 \cite{EN50128:2011RAMS} used in railway applications.

Our SM draws on 26 studies, selected out of 4,751 using a systematic process composed of several stages.
All 26 studies address formal and/or semi-formal languages and/or techniques.
An important feature of our SM is that, although we focused on the SoS domain, we did not narrow our search by including SoS in our search string, which allowed us to search more deeply for information regarding the state of the art in formal and semi-formal modeling of SoS requirements.
We found 15 studies modeling part or all SoS requirements using only formal languages or techniques, five adopting only semi-formal ones, and another six combining formal and semi-formal ones.

SoS have an inherent complexity in their development with new requirements emerging at runtime and a multiplicity of constituents, which are sometimes unknown at design time or can change at runtime.
Modeling the requirements of these SoS and their constituents certainly requires a combination of formal and/or semi-formal languages and techniques. 
The adoption of formal modeling in all systems requirements may increase the development cost and the time spent on specification, but could minimize errors introduced in other lifecycles phases. 
Besides that, formal languages and techniques offer the precision necessary for modeling the requirements of critical, complex systems; however, they require mathematical expertise from their users. A common strategy is to encapsulate the formal information in specification tools, which minimizes the need by for mathematical expertise, maintains the precision of the requirements specifications, and could encourage adoption the industry.
On the order hand, modeling using only semi-formal languages and techniques may reduce the time needed to produced a specification, but might increase the number of errors introduced in the development process. 

We identified 32 formal and semi-formal techniques and languages, of which 19 are executable and one language (i.e., XML) is interpreted. In other words, requirements specified in these 20 techniques/languages can be evaluated automatically with tool support. 
Tools can increase the precision of analyses and the accuracy of the completeness, consistency, and correctness of SoS requirements.
Moreover, 25 languages/techniques from the 38 one found are well documented (cf. Table \ref{tab:ClassificationStyle}, which might facilitate their adoption. 
However, the research field is relatively new to SoS and 
some studies from other domains, like embedded systems, cyber-physical systems, and self-adaptive systems, were found that propose ideas     that may be adapted to SoS.

A deeper analysis of formal and semi-formal V\&V of SoS requirements still needs to be performed. 
We also observed that most of these approaches have been applied to systems with characteristics compatible with SoS, but not exactly SoS, considering their inherent characteristics. 
Our investigation only revealed three (S21, S23, and S426) where these characteristics (i.e., managerial and/or operational independence, evolutionary development, emergent behavior, and geographical distribution) are addressed explicitly.
Hence, we believe that more attention still needs to be paid to SoS RE, specifically regarding modeling and verification, which are activities that ensure high-quality requirements. 
Finally, tools that consider not only requirements at the SoS level, but also the elicitation, modeling, and verification of the constituents' requirements need to be developed in a more integrated way.


\section{Conclusions} \label{conclusion}

Formal and semi-formal modeling of SoS requirements can certainly contribute to improvinge the quality of these systems, mainly when applied together with requirements verification. 
Even considering the advantages of formal modeling for critical systems development, the use of formal modeling in the industrial sector is still hindered by two specifics issues: 
(i) the cost of formal application is high; and 
(ii) specialized professionals are required to understand the semantics of specification languages.
In this scenario, the main contribution of this work is a landscape of languages and techniques for formally and semi-formally modeling SoS requirements, including initiatives adopted in similar systems that could be useful for SoS. 
For this purpose, we applied a systematic approach to the identification and analysis of the studies. 

As future work (besides of the needs already mentioned) we intend to perform: 
(i) a more specific investigation in this research area, for instance, the way that formal and semi-formal V\&V of SoS requirements have been addressed and integrated with modeling; and 
(ii) consolidation of the results of this SM, aimed at providing a more detailed analysis of all the evidence presented in this work. 
Finally, this work is intended to direct the attention of researchers and practitioners to the importance of adequately treating requirements modeling, particularly when such systems are as complex, critical, and software-intensive as SoS.


\ifCLASSOPTIONcompsoc
  \section*{Acknowledgments}
\else
  \section*{Acknowledgment}
\fi

This work is supported by the S\~ao Paulo Research Foundation (FAPESP) under grant no.{2015/06195-3}, no.{2017/15354-3}, and  no.{2017/06195-9}.
We also thank the SofTware Architecture Research Team (START) of
ICMC/USP, Brazil, for its advice and guidance in improvinge this work, the anonymous reviewers for their thorough reviews, and Sonnhild Namingha from the Fraunhofer (IESE), Kaiserslautern, Germany, for precious advice and language review.

\ifCLASSOPTIONcaptionsoff
  \newpage
\fi



%
\bibliographystyle{IEEEtran}
\bibliography{2017-12-27-lana-references_CompWith26}

%




\end{document}